\documentclass{ws-procs975x65}

\begin{document}

\title{Neutron star versus heavy-ion data: is the nuclear equation
  of state hard or soft?}

\author{Irina Sagert, Mirjam Wietoska, J\"urgen Schaffner-Bielich$^*$}

\address{Institut f\"ur Theoretische Physik/Astrophysik, 
J. W. Goethe Universit\"at,\\ 
D-60438 Frankfurt am Main, Germany\\
$^*$E-mail: schaffner@astro.uni-frankfurt.de}

\author{Christian Sturm}

\address{Institut f\"ur Kernphysik, 
J. W. Goethe Universit\"at,\\ 
D-60438 Frankfurt am Main, Germany}

\begin{abstract}
  Recent astrophysical observations of neutron stars and heavy-ion data
  are confronted with our present understanding of the equation of state
  of dense hadronic matter. Emphasis is put on the possible role of the
  presence of hyperons in the interior of compact stars. We argue that
  data from low-mass pulsars provide an important cross-check between
  high-density astrophysics and heavy-ion physics.
\end{abstract}

\keywords{nuclear equation of state; neutron stars; pulsars; kaon
  production in heavy-ion collisions}

\bodymatter

\section{Introduction}

The research areas of high-density astrophysics, as the physics of
compact stars, and relativistic heavy-ion collisions are probing matter
at extreme densities.  The properties of neutron stars are determined by
the nuclear equation of state (EoS), as well as microphysical reactions in
dense matter. The stiffness of the high-density matter controls the
maximum mass of compact stars. New measurements of the global properties
of pulsars, rotation-powered neutron stars, point towards large masses
and correspondingly to a rather stiff equation of state (for a recent
review on the equation of state for compact stars see
\cite{Lattimer:2006xb}). In a recent analysis of the x-ray burster EXO
0748--67 it was even claimed that soft nuclear equations of state are
ruled out \cite{Ozel:2006km}. Note that this analysis, if confirmed,
would not rule out the presence of quark matter in the core of compact
stars \cite{Alford:2006vz}.

On the other side, strange particles (kaons) produced in relativistic
heavy-ion collisions just below the threshold of the elementary reaction
are sensitive to medium effects due to the created high-density matter
(see e.g.\ \cite{Schaffner:1996kv}). Recent investigations conclude that
the systematics of kaon production can only be explained by an extremely
soft nuclear equation of state above normal nuclear matter density
\cite{Sturm:2000dm,Fuchs:2000kp,Hartnack:2005tr,Forster:2007qk}.

There seems to be conflict in determining the nuclear equation
of state, which we will discuss in detail in the following. We
investigate the impacts of the compression modulus and symmetry energy
of nuclear matter on the maximum mass of neutron stars in view of the
recent constraints from heavy-ion data on kaon production in dense
matter. In particular, we delineate the different density regions probed
in the mass-radius diagram of compact stars. We outline the importance
of the Schr\"odinger equivalent potentials for subthreshold production
of kaons. The possible effects from the presence of hyperons in dense
neutron star matter are confronted with pulsar mass measurements.

\section{The nuclear EoS from astrophysical and 
heavy-ion data: a soft or hard EoS?}

The properties of high-density nuclear matter is intimately related to
the phase diagram of quantum chromodynamics (QCD), for a review see
e.g.\ \cite{Rischke:2003mt}. The regime of high temperatures and nearly
vanishing baryochemical potential is probed by present and ongoing
heavy-ion experiments at BNL's Relativistic Heavy-Ion Collider and
CERN's Large Hadron Collider and is related to the physics of the early
universe. A rapid crossover transition due to chiral symmetry
restoration and deconfinement is found in lattice gauge simulations, see
e.g.\ \cite{Cheng:2007jq}. The QCD phase diagram at large baryochemical
potential and moderate temperatures constitutes the region of the chiral
phase transition and the high-density astrophysics of core-collapse
supernovae and compact stars (see e.g.\ \cite{SchaffnerBielich:2007mr}
for a recent treatise). Terrestrial heavy-ion experiments, as the
Compressed Baryonic Matter (CBM) experiment at GSI's Facility for
Antiproton and Ion Research (FAIR) will investigate this fascinating and
largely unknown terrain of the QCD phase diagram \cite{Senger:2006zz}.

The nuclear equation of state serves as a crucial input for simulations
of core-collapse supernovae \cite{Janka:2006fh}, neutron star mergers
\cite{Rosswog:2001fh,Oechslin:2006uk}, proto-neutron star evolution
\cite{Pons:1998mm} and, of course for determining the properties of cold
neutron stars \cite{Weber:2004kj}. Pulsar mass measurements provide
constraints on the stiffness of the nuclear equation of state.
Unfortunately, out of the more than 1600 known pulsars, only a few
precise mass measurements from binary pulsars are currently available
(see \cite{Stairs:2006yr} and references therein). Still, the
undoubtedly upper mass limit is given by the Hulse-Taylor pulsar of
$M=(1.4414\pm 0.0002)M_\odot$ \cite{Weisberg:2004hi}, the lightest
pulsar known is J1756-2251 with a mass of $M=(1.18\pm 0.02)M_\odot$
\cite{Faulkner:2005}. New data on pulsar masses has been presented at
the Montreal conference on pulsars (see ns2007.org). The mass of the
pulsar J0751+1807, originally with a median above two solar masses with
$M=2.1\pm 0.2 M_\odot$ ($1\sigma$) \cite{Nice:2005fi}, is now corrected
and below the Hulse-Taylor mass limit \cite{Stairs2007}. However, the
mass of the pulsar J0621+1002 was determined to be between 1.53 to 1.80
solar masses ($2\sigma$) \cite{Kasian2007}. Combined data from the
pulsars Terzan 5I and J \cite{Ransom:2005ae} with the pulsar B1516+02B
\cite{Freire:2007ux} results in a mass limit of 1.77 solar masses for at
least one of these pulsars. Measurement of the pulsar J1748--2021B
arrives at a lower mass limit of $M>2M_\odot$ \cite{Freire:2007ux} but
that could be the mass of a two neutron star system.  The analysis of
x-ray burster is much more model dependent. For EXO 0748--676 a
mass-radius constraint of $M\geq 2.10\pm 0.28 M_\odot$ and $R\geq 13.8
\pm 1.8$ km has been derived \cite{Ozel:2006km}, for a critical
discussion on the analysis I refer to \cite{Walter2006}. That constraint
would actually rule out soft nuclear equations of state but not the
presence of quark matter, as quark matter is an entirely new phase which
can be rather stiff \cite{Alford:2006vz}.

High-density nuclear matter is produced in the laboratory for a fleeting
moment of time in the collisions of heavy nuclei at relativistic
bombarding energies. The properties of kaons can change substantially in
the high-density matter created. The in-medium energy of kaons will
increase with density (basically due to the low-density theorem, see
however \cite{Korpa:2004ae} which arrives at a somewhat stronger
repulsive potential). Kaons are produced by the associated production
mechanism NN$\to{\rm N}\Lambda$K, NN$\to$NNK$\overline{\rm K}$, and most
importantly by the in-medium processes $\pi{\rm N}\to\Lambda$K,
$\pi\Lambda\to{\rm N}\overline{\rm K}$, which are rescattering processes
of already produced particles.  The effective energy of kaons in the
medium will change the Q-values of the direct production and
rescattering processes, therefore affecting the net production rate
\cite{Schaffner:1996kv}.  As kaons have long mean free paths, they can
leave the high-density region and serve as an excellent tool to probe
its properties.  Indeed, detailed transport simulations find that
nuclear matter is compressed up to $3n_0$ for a typical bombarding
energy of 1 to 1.5~AGeV and that the produced kaons are dominantly produced
around $2 n_0$, where $n_0$ stands for the normal nuclear matter
saturation density \cite{Fuchs:2000kp,Hartnack:2005tr}. Kaons are
produced below the elementary threshold energy due to multi-step
processes which increase with the maximum density achieved in the
collisions. The double ratio of the multiplicity per mass number for the
C+C collisions and Au+Au collisions turns out to be rather insensitive
to the input parameters (elementary cross sections, in-medium potential)
which scale linearly with mass number or density. Only calculations with
a compression modulus of $K\approx 200$ MeV can describe the trend of
the kaon production data
\cite{Sturm:2000dm,Fuchs:2000kp,Hartnack:2005tr,Forster:2007qk}. Hence,
the analysis of heavy-ion experiments points towards a rather soft
nuclear equation of state.

\section{The different density regimes of neutron stars}

In the following we discuss the different densities encountered in
neutron stars and the corresponding regions in the mass-radius diagram.
While the standard lore is that the crust of a neutron star consists of
nuclei, neutrons and electrons, the composition of the interior of a
neutron star is basically unknown. At about $2n_0$ hyperons can appear
as a new hadronic degree of freedom. Kaons can be formed as Bose-Einstein
condensate. Finally, chirally restored quark matter can be present as an
entirely new phase in the core of compact stars. After considering pure
nucleonic matter, we focus on the role of hyperons and their
importance for the properties of neutron stars (see also
\cite{SchaffnerBielich:2007tj} and references therein).

First, let us consider just nucleonic matter. Its equation of state can
be modelled by a Skyrme-type ansatz for the energy per nucleon. The
parameters are fixed by the nuclear matter properties, as the saturation
density, binding energy, compression modulus and asymmetry energy
\cite{Sagert:2007nt}. In addition, we explore effects from the asymmetry
term having a density dependence which scales with a power $\alpha$ as
extracted from heavy-ion collision measurements where $\alpha$ is
between 0.7 and 1.1 \cite{Chen:2004si,Li:2005sr,Chen:2007fsa}. The
pressure is determined by a thermodynamic relation, which fixes completely
the EoS used in transport simulations of heavy-ion collisions. With that
EoS at hand one can check now, whether the low compressibilities found
in describing the kaon production data of $K \approx 200$~MeV
\cite{Sturm:2000dm,Fuchs:2000kp,Hartnack:2005tr,Forster:2007qk} are
ruled out by neutron star measurements.

Solving the Tolman-Oppenheimer-Volkoff equation gives the result that
rather large maximum neutron star masses can be reached even for such
low values of the compression modulus \cite{Sagert:2007nt}. The maximum
mass is greater than $M>2M_\odot$ for a compression modulus of
$K>160$~MeV ($\alpha=1.0$, 1.1) and for the case $\alpha=0.7$ greater
than $M> 1.6M_\odot$ for $K>160$~MeV and greater than $M> 2M_\odot$ for
$K>220$~MeV. Changing the asymmetry energy within reasonable values
($S_0=28$ to 32~MeV) shifts the maximum mass by at most $\Delta M=\pm
0.1M_\odot$ for low values of $K$.  The maximum central density is
about $n_c=(7\div 8)n_0$ for $\alpha=1.0,1.1$ and can hit even $10n_0$ for
$\alpha=0.7$. The EoS is causal up to a compression modulus of $K=340$,
corresponding to a maximum mass of $M=2.6M_\odot$, for $\alpha=1.0,1.1$
and up to $K=280$ MeV for $\alpha=0.7$. Hence, we conclude that even a
pulsar mass of $2M_\odot$ would be compatible with the 'soft' EoS as
extracted from heavy-ion data. This statement is corroborated by more
advanced many-body approaches to the nuclear EoS for kaon production in
heavy-ion collisions and neutron star mass limits \cite{Fuchs:2007vt}.

\begin{figure}[t]
\begin{center}
\includegraphics[width=0.8\textwidth]{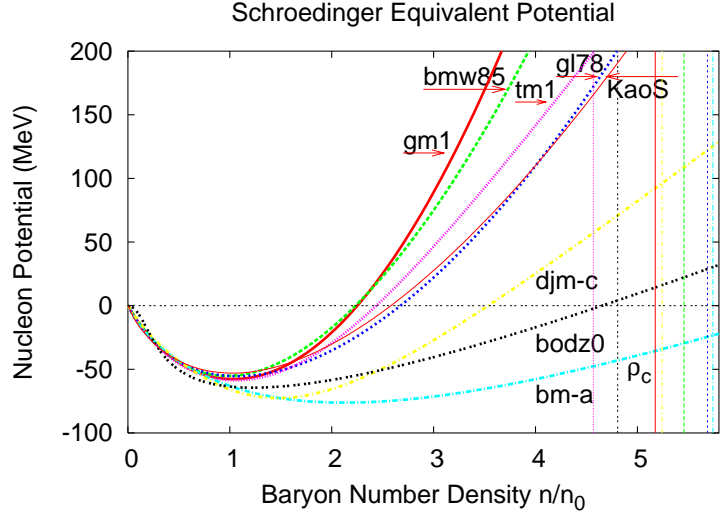}
\end{center}
\caption{The Schr\"odinger equivalent potential versus the baryon number
  density given in $n_0$ for various parameters of the relativistic
  mean-field model.}
\label{fig:sevpot}
\end{figure}

For a field-theoretical investigation on the nuclear equation of state
in heavy-ion collisions and for neutron stars one has to consider the
Schr\"odinger equivalent potential, which is the actual input to the
transport simulation codes, not the nuclear equation of state. There is
a one-to-one correspondence between the energy per baryon and the
nucleon potential for the non-relativistic Skyrme model as studied
above.  However, this direct relation is lost in a relativistic field
theoretical approach as for example the relativistic nucleon potential
exhibits now a scalar and vector part. We note also, that the direct
relation between the compression modulus $K$ and the stiffness of the
nuclear equation of state at supra-nuclear densities is also lost. The
stiffness of the EoS in the standard relativistic mean-field (RMF) model
is controlled by the effective mass of the nucleon at saturation density
not by the compression modulus, which is actually well known for quite
some time, see e.g.\ \cite{Waldhauser:1987ed}. The Schr\"odinger
equivalent potential for a sample of parameter sets of the relativistic
mean-field model is depicted in Fig.~\ref{fig:sevpot}. The line marked
'KaoS' stands for the nucleon potential as used in transport simulations
for a compression modulus of $K=200$~MeV. In order to be in accord with
the KaoS data, the potential of the relativistic mean-field parameter
set should be below the curve labelled 'KaoS' at a density region of
around $2n_0$ where most of the kaons are produced at subthreshold
collision energies. The parameter sets used for the standard nonlinear
RMF model are 'bmw85' with an effective mass of $m^*/m=0.85$ and
$K=300$~MeV \cite{Waldhauser:1987ed}, 'gm1' with $m^*/m=0.7$ and
$K=300$~MeV, and 'gl78' with $m^*/m=0.78$ and $K=240$~MeV
\cite{Glendenning:1991es}. Note, that the values chosen for the
effective nucleon mass are quite high so that the nuclear EoS becomes
soft.  Typical fits to properties of nuclei arrive at values of
$m^*/m\approx 0.6$ as for the parameter set 'tm1' which is fitted to
properties of spherical nuclei \cite{Sugahara:1993wz}. The set 'tm1' has
an additional selfinteraction term for the vector fields which results
in an overall similar behaviour of the nucleon potential in comparison
to the other RMF parameter sets with a soft EoS.  Those vector
selfinteractions were introduced in \cite{Bodmer:1991hz} where the set
'bodz0' with $m^*/m=0.6$ and $K=300$~MeV is taken from.  One motivation
of introducing this vector selfinteraction term is to describe the
nucleon vector potential as computed in more advanced many-body
approaches which are based on nucleon-nucleon potentials. For the sets
'bm-a' and 'djm-c', the vector selfenergy of the nucleon in the RMF
calculation was adjusted to the ones of Dirac-Brueckner-Hartree-Fock
calculations \cite{Gmuca:1992}. The minima of the nucleon potential of
those latter three parameter sets are located at larger densities than
the saturation density, in particular for the set 'bm-a'. We stress that
the nuclear equation of state for those sets, however, gives the right
properties of saturated nuclear matter \cite{Gmuca:1992}.
Fig.~\ref{fig:sevpot} shows also the maximum density reached in the
center of the maximum mass configuration of the neutron star sequence by
vertical lines, which are surprisingly close lined up between 4.5 to
$6n_0$ in view of the large differences in the nucleon potential at high
densities.
 
\begin{figure}[t]
\begin{center}
\includegraphics[width=0.75\textwidth]{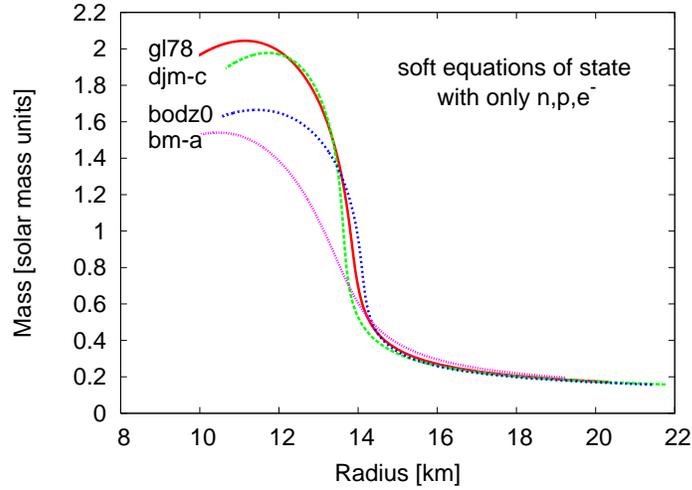}
\end{center}
\caption{The mass-radius plot for various parameter sets
  of the relativistic mean-field model with nucleons and leptons only.}
\label{fig:mrnpe}
\end{figure}

The mass-radius diagram for the RMF parameter sets giving a small
Schr\"odinger equivalent potential, i.e.\ one which is at or below the
potential used in transport simulations ($K=200$~MeV), is plotted in
Fig.~\ref{fig:mrnpe} for neutron star matter consisting of nucleons and
leptons only. The sets 'gl78' and 'djm-c' reach maximum masses of
$2.04M_\odot$ and $1.98M_\odot$, respectively, even though the nucleon
potential for the set 'djm-c' is well below the limit given from the
heavy-ion data analysis (see Fig.~\ref{fig:sevpot}). The sets 'bodz0'
and 'bm-a' just arrive at maximum masses of $1.66M_\odot$ and
$1.54M_\odot$, respectively, which could be ruled out with a confirmed
measurement of a heavy pulsar but so far can not be excluded. It seems,
that the constraint from heavy-ion data on the nucleon potential alone
is in agreement with pulsar mass measurements for relativistic
mean-field approaches, even if masses of about $2M_\odot$ will be
measured in the future. An important point to stress here is that the
heavy-ion data and the determination of the maximum mass of neutron
stars addresses completely different density regimes. While the
heavy-ion data on kaon production probes at maximum 2 to $3n_0$, the
central density of the most massive neutron stars tops $5n_0$.  Hence,
the maximum mass of neutron stars probes the high-density regime of the
nuclear equation of state which is not constrained by the heavy-ion data
presently available. In other words, if the pressure, or better the
nucleon potential, rises slowly at densities up to $2n_0$, it could
increase rapidly at larger densities so as to comply with astrophysical
data on neutron star masses. Moreover, new particles and phases could
certainly appear at such large densities which change the equation of
state for massive neutron stars substantially, as hyperon matter, to
which we turn now for making our argumentation more explicit.

\begin{figure}[t]
\begin{center}
\includegraphics[width=0.75\textwidth]{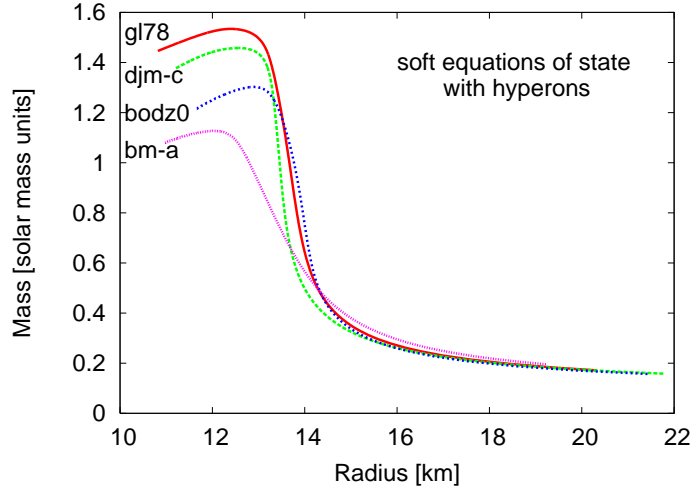}
\end{center}
\caption{The mass-radius plot for various parameter sets of the relativistic
  mean-field model including the effect of hyperons.}
\label{fig:mrnpehyp}
\end{figure}

The in-medium properties of hyperons are constrained by hypernuclear
data. In particular, the $\Lambda$ potential at $n_0$ is quite well
determined to be $-30$~MeV. Other hyperon potential are much less well
known, unfortunately. Hyperons, if present, have a strong impact on the
properties of compact stars (see \cite{SchaffnerBielich:2007tj} for a
recent outline). $\Lambda$ hyperons constitute a new hadronic degree of
freedom in neutron star matter at and above about $2n_0$. The population
of other hyperons, $\Sigma$ and $\Xi$ hyperons, is highly sensitive to
their in-medium potentials. For a slightly repulsive potential for the
$\Sigma$ these hyperons do not appear in compact star matter at all. The
$\Xi$ hyperons will be present, as their in-medium potential is likely to
be attractive. The presence of hyperons changes drastically the
properties of compact stars. The new degree of freedom lowers the
pressure for a given energy density, so that the EoS is considerably
softened at large densities. There is a substantial decrease of the
maximum mass due to the appearance of hyperons in compact stars: the
maximum mass for such ``giant hypernuclei'' can drop down by $\Delta
M\approx 0.7 M_\odot$ compared to the case of neutron star matter
consisting of nucleons and leptons only \cite{Glendenning:1991es}. For
the RMF parameter sets studied here, we add hyperons as outlined in
\cite{Schaffner:1995th} by fixing the hyperon vector coupling constants
via SU(6) symmetry relations and the hyperon scalar coupling constant to
the (relativistic) hyperon potentials as determined in
\cite{SchaffnerBielich:2000wj} from hypernuclear data and hyperonic
atoms. The resulting mass-radius plot when including hyperons is
pictured in Fig.~\ref{fig:mrnpehyp}. Note the different mass scales of
Figs.~\ref{fig:mrnpe} and \ref{fig:mrnpehyp}, the maximum mass with
hyperons included is now substantially decreased to $1.53 M_\odot$ for
the set 'gl78', to $1.46 M_\odot$ for the set 'djm-c', to $1.30 M_\odot$
for the set 'bodz0', and to $1.27 M_\odot$ for the set 'bm-a'. The
latter two cases are now even below the Hulse-Taylor mass limit and can
be ruled out. The former two cases are just above the Hulse-Taylor mass
limit of $1.44M_\odot$ and could be ruled out if measurements of heavy
neutron stars masses of $1.6M_\odot$ or more will be confirmed in future
astrophysical observations. Clearly, the presence of hyperons in compact
stars could be severely constrained by combining the heavy-ion data
analysis with the measurement of a heavy neutron star. The limit on the
nucleon potential from heavy-ion data seems to make it quite difficult
to reach neutron star masses above say $1.6M_\odot$ for the RMF model
when hyperons are included via SU(6) symmetry and by adopting the
presently (sometimes poorly) known hyperon potentials from hypernuclear
data.  Of course, a much more systematic analysis needs to be done, a
firm statement can not be drawn from our sample of parameter sets.
Certainly, the situation could be changed for other many-body
approaches. Within the relativistic Hartree-Fock approach, for example,
maximum masses of about $1.9M_\odot$ are possible even when effects from
hyperons are added to the equation of state \cite{Huber:1997mg}. But one
rather robust conclusion can be drawn from our analysis: the
high-density EoS above $2n_0$, where hyperons appear and modify the EoS
in the models used here, is crucial in determining the maximum mass of a
neutron star.  Hence, we are probing this density region when looking at
the maximum mass configurations of compact stars.

The procedure to follow is now eminent, the true comparison between the
present heavy-ion data and astrophysical data on compact stars is not
located in the high-mass region but on the low-mass region of the
mass-radius diagram of compact stars. The lightest neutron star known at
present is the pulsar J1756-2251 with a mass of $M=(1.18\pm 0.02)M_\odot$
\cite{Faulkner:2005}. Much lower values are probably not realized in
nature as hot proto-neutron stars have a much larger minimum stable mass
than cold neutron stars, for example a minimum mass of $0.86M_\odot$ has
been found for an isothermal proto-neutron star \cite{Gondek:1997}.
Interestingly, a $1.2M_\odot$ neutron star has a maximum density of
$n=2n_0$ in our non-relativistic models \cite{Sagert:2007nt}, so that
exotic matter is likely to be not present.  We find that the radius of
such a low-mass neutron star is in fact highly sensitive to the nuclear
equation of state (see also \cite{Lattimer:2000nx}), in particular to
the asymmetry energy at high densities which is well known
\cite{Horowitz:2000xj,Horowitz:2001ya,Steiner:2004fi,Li:2005sr}. There
are several promising proposals for radii measurements of neutron stars,
see \cite{Lattimer:2006xb} for a recent overview. The fascinating
aspect is that heavy-ion experiments can address this density region and
probe not only the equation of state but also the density dependence of
the asymmetry energy. The ratio of the produced isospin partners K$^+$
and K$^0$ at subthreshold energies has been demonstrated to be sensitive
to the isovector potential above saturation density
\cite{Ferini:2006je}. The tantalising conclusion is that a direct
comparison with heavy-ion data and compact star data seems to be
feasible.

As always there are exceptions to the assumption, that the nuclear EoS
just contains nucleons and leptons up to $1.2M_\odot$. In
Ref.~\cite{Schertler:2000xq} strange quark matter is already present for
only $0.3M_\odot$ which depends hugely on the choice of the MIT bag
constant. In Ref.~\cite{Schulze:2006vw} hyperons appear already for a
compact star mass of only $0.5M_\odot$ although the critical
density for the onset of the hyperon population is around $2n_0$. The
reason is that the equation of state is unphysically soft, so that the
maximum mass is below the Hulse-Taylor mass limit. In any case, this
provides another opportunity for the radius measurement of low-mass
pulsars: if their radii turn out to be completely off the range
predicted from our knowledge of the density dependence of the asymmetry
energy, some exotic matter is present in the core of neutron stars!

\section{Summary}

The combined analysis of heavy-ion data on kaon production at
subthreshold energies and neutron star mass measurements points towards
a nuclear EoS that is soft at moderate densities and hard at high
densities. A soft nuclear EoS as extracted from kaon production data is
not in contradiction with heavy pulsars as mutually exclusive density
regions are probed. The nuclear EoS above $n\approx 2n_0$ determines the
maximum mass of neutron stars, which is controlled by unknown
high-density physics (as hyperons and quark matter). Compact star matter
constrained by the heavy-ion data seems to result in rather low maximum
masses for compact stars when hyperons are included. A measurement of a
heavy pulsar will make it quite difficult for having hyperons inside a
neutron star and could exhibit an emerging conflict between hypernuclear
and pulsar data.  Properties of low-mass neutron stars ($M\leq
1.2M_\odot$), however, are likely to be entirely determined by the EoS
of nucleons and leptons only up to $n\approx 2n_0$, as hyperon and
possibly quark matter could appear at larger densities. Thus, the
measurement of the radii of low-mass pulsars provides the opportunity
for a cross-check between heavy-ion and astrophysical data and possibly
for the detection of an exotic phase in the interior of compact stars.

\section*{Acknowledgements}

This work is supported in part by the Gesellschaft f\"ur
Schwerionenforschung mbH, Darmstadt, Germany. Irina Sagert gratefully
acknowledges support from the Helmholtz Research School for Quark Matter
Studies.

\bibliographystyle{ws-procs975x65}
\bibliography{all,literat,exoct2007}

\begin{thebibliography}{10}

\bibitem{Lattimer:2006xb}
J.~M. Lattimer and M.~Prakash, {\em Phys. Rept.} {\bf 442}, 109 (2007).

\bibitem{Ozel:2006km}
F.~{\"O}zel, {\em Nature} {\bf 441}, 1115 (2006).

\bibitem{Alford:2006vz}
M.~Alford, D.~Blaschke, A.~Drago, T.~Kl\"ahn, G.~Pagliara and
  J.~Schaffner-Bielich, {\em Nature} {\bf 445}, E7 (2006).

\bibitem{Schaffner:1996kv}
J.~Schaffner-Bielich, I.~N. Mishustin and J.~Bondorf, {\em Nucl. Phys.} {\bf
  A625}, p. 325 (1997).

\bibitem{Sturm:2000dm}
C.~Sturm {\em et~al.}, {\em Phys. Rev. Lett.} {\bf 86}, 39 (2001).

\bibitem{Fuchs:2000kp}
C.~Fuchs, A.~Faessler, E.~Zabrodin and Y.-M. Zheng, {\em Phys. Rev. Lett.} {\bf
  86}, 1974 (2001).

\bibitem{Hartnack:2005tr}
C.~Hartnack, H.~Oeschler and J.~Aichelin, {\em Phys. Rev. Lett.} {\bf 96}, p.
  012302 (2006).

\bibitem{Forster:2007qk}
A.~Forster {\em et~al.}, {\em Phys. Rev. C} {\bf 75}, p. 024906 (2007).

\bibitem{Rischke:2003mt}
D.~H. Rischke, {\em Prog. Part. Nucl. Phys.} {\bf 52}, 197 (2004).

\bibitem{Cheng:2007jq}
M.~Cheng, N.~H. Christ, S.~Datta, J.~van~der Heide, C.~Jung, F.~Karsch,
  O.~Kaczmarek, E.~Laermann, R.~D. Mawhinney, C.~Miao, P.~Petreczky, K.~Petrov,
  C.~Schmidt, W.~Soeldner and T.~Umeda, {\em arXiv:0710.0354 [hep-lat]}
  (2007).

\bibitem{SchaffnerBielich:2007mr}
J.~Schaffner-Bielich, {\em arXiv:0709.1043 [astro-ph]}  (2007).

\bibitem{Senger:2006zz}
P.~Senger, T.~Galatyuk, D.~Kresan, A.~Kiseleva and E.~Kryshen, {\em PoS} {\bf
  CPOD2006}, p. 018 (2006).

\bibitem{Janka:2006fh}
H.-T. Janka, K.~Langanke, A.~Marek, G.~Mart{\'i}nez-Pinedo and B.~M{\"u}ller,
  {\em Phys. Rept.} {\bf 442}, 38 (2007).

\bibitem{Rosswog:2001fh}
S.~Rosswog and M.~B. Davies, {\em Mon. Not. Roy. Astron. Soc.} {\bf 345}, p.
  1077 (2003).

\bibitem{Oechslin:2006uk}
R.~{Oechslin}, H.-T. {Janka} and A.~{Marek}, {\em Astron. Astrophys.} {\bf
  467}, 395 (2007).

\bibitem{Pons:1998mm}
J.~A. Pons, S.~Reddy, M.~Prakash, J.~M. Lattimer and J.~A. Miralles, {\em
  Astrophys. J.} {\bf 513}, p. 780 (1999).

\bibitem{Weber:2004kj}
F.~Weber, {\em Prog. Part. Nucl. Phys.} {\bf 54}, 193 (2005).

\bibitem{Stairs:2006yr}
I.~H. Stairs, {\em J. Phys. G} {\bf 32}, S259 (2006).

\bibitem{Weisberg:2004hi}
J.~M. Weisberg and J.~H. Taylor, The relativistic binary pulsar b1913+16:
  Thirty years of observations and analysis, in {\em Binary Radio Pulsars\/},
  eds. F.~A. Rasio and I.~H. Stairs, Astronomical Society of the Pacific
  Conference Series, Vol.~328 2005 p.~25.

\bibitem{Faulkner:2005}
A.~J. {Faulkner}, M.~{Kramer}, A.~G. {Lyne}, R.~N. {Manchester}, M.~A.
  {McLaughlin}, I.~H. {Stairs}, G.~{Hobbs}, A.~{Possenti}, D.~R. {Lorimer},
  N.~{D'Amico}, F.~{Camilo} and M.~{Burgay}, {\em Astrophys. J.} {\bf 618},
  L119 (2005).

\bibitem{Nice:2005fi}
D.~J. Nice, E.~M. Splaver, I.~H. Stairs, O.~L\"ohmer, A.~Jessner, M.~Kramer and
  J.~M. Cordes, {\em Astrophys. J.} {\bf 634}, 1242 (2005).

\bibitem{Stairs2007}
David Nice and Ingrid Stairs, private communication.

\bibitem{Kasian2007}
Laura Kasian, David Nice and Ingrid Stairs, private communication.

\bibitem{Ransom:2005ae}
S.~M. Ransom, J.~W.~T. Hessels, I.~H. Stairs, P.~C.~C. Freire, F.~Camilo, V.~M.
  Kaspi and D.~L. Kaplan, {\em Science} {\bf 307}, 892 (2005).

\bibitem{Freire:2007ux}
P.~C.~C. Freire, S.~M. Ransom, S.~Begin, I.~H. Stairs, J.~W.~T. Hessels, L.~H.
  Frey and F.~Camilo, {\em arXiv:0711.2028 [astro-ph]}  (2007).

\bibitem{Walter2006}
F.~M. {Walter} and J.~M. {Lattimer}, {\em Nature Physics} {\bf 2}, 443 (2006).

\bibitem{Korpa:2004ae}
C.~L. Korpa and M.~F.~M. Lutz, {\em Acta Phys. Hung.} {\bf A22}, 21 (2005).

\bibitem{SchaffnerBielich:2007tj}
J.~Schaffner-Bielich, {\em astro-ph/0703113}  (2007).

\bibitem{Sagert:2007nt}
I.~Sagert, M.~Wietoska, J.~Schaffner-Bielich and C.~Sturm, {\em arXiv:0708.2810
  [astro-ph]}  (2007).

\bibitem{Chen:2004si}
L.-W. Chen, C.~M. Ko and B.-A. Li, {\em Phys. Rev. Lett.} {\bf 94}, p. 032701
  (2005).

\bibitem{Li:2005sr}
B.-A. Li and A.~W. Steiner, {\em Phys. Lett.} {\bf B642}, 436 (2006).

\bibitem{Chen:2007fsa}
L.-W. Chen, C.~M. Ko, B.-A. Li and G.-C. Yong, {\em arXiv:0704.2340 [nucl-th]}
  (2007).

\bibitem{Fuchs:2007vt}
C.~Fuchs, {\em arXiv:0706.0130 [nucl-th]}  (2007).

\bibitem{Waldhauser:1987ed}
B.~M. Waldhauser, J.~A. Maruhn, H.~St{\"o}cker and W.~Greiner, {\em Phys. Rev.
  C} {\bf 38}, 1003 (1988).

\bibitem{Glendenning:1991es}
N.~K. Glendenning and S.~A. Moszkowski, {\em Phys. Rev. Lett.} {\bf 67}, p.
  2414 (1991).

\bibitem{Sugahara:1993wz}
Y.~Sugahara and H.~Toki, {\em Nucl. Phys.} {\bf A579}, 557 (1994).

\bibitem{Bodmer:1991hz}
A.~R. Bodmer, {\em Nucl. Phys.} {\bf A526}, 703 (1991).

\bibitem{Gmuca:1992}
S.~Gmuca, {\em Nucl. Phys.} {\bf A547}, 447 (1992).

\bibitem{Schaffner:1995th}
J.~Schaffner and I.~N. Mishustin, {\em Phys. Rev. C} {\bf 53}, p. 1416 (1996).

\bibitem{SchaffnerBielich:2000wj}
J.~Schaffner-Bielich and A.~Gal, {\em Phys. Rev. C} {\bf 62}, p. 034311 (2000).

\bibitem{Huber:1997mg}
H.~Huber, F.~Weber, M.~K. Weigel and C.~Schaab, {\em Int. J. Mod. Phys.} {\bf
  E7}, 301 (1998).

\bibitem{Gondek:1997}
D.~Gondek, P.~Haensel and J.~L. Zdunik, {\em Astron. Astrophys.} {\bf 325}, 217
  (1997).

\bibitem{Lattimer:2000nx}
J.~M. Lattimer and M.~Prakash, {\em Astrophys. J.} {\bf 550}, p. 426 (2001).

\bibitem{Horowitz:2000xj}
C.~J. Horowitz and J.~Piekarewicz, {\em Phys. Rev. Lett.} {\bf 86}, p. 5647
  (2001).

\bibitem{Horowitz:2001ya}
C.~J. Horowitz and J.~Piekarewicz, {\em Phys. Rev.} {\bf C64}, p. 062802
  (2001).

\bibitem{Steiner:2004fi}
A.~W. Steiner, M.~Prakash, J.~M. Lattimer and P.~J. Ellis, {\em Phys. Rept.}
  {\bf 411}, 325 (2005).

\bibitem{Ferini:2006je}
G.~Ferini, T.~Gaitanos, M.~Colonna, M.~Di~Toro and H.~H. Wolter, {\em Phys.
  Rev. Lett.} {\bf 97}, p. 202301 (2006).

\bibitem{Schertler:2000xq}
K.~Schertler, C.~Greiner, J.~Schaffner-Bielich and M.~H. Thoma, {\em Nucl.
  Phys.} {\bf A677}, p. 463 (2000).

\bibitem{Schulze:2006vw}
H.~J. Schulze, A.~Polls, A.~Ramos and I.~Vidana, {\em Phys. Rev. C} {\bf 73},
  p. 058801 (2006).

\end{thebibliography}

\end{document}